\documentclass{ws-p8-50x6-00}

\def\gamgam{\mbox{$\gamma \gamma$}}
\def\gamp{\mbox{$\gamma \ p$}}

\begin{document}
\title{Total Cross-sections}

\author{Rohini M.\ Godbole}
\address{ Centre for Theoretical Studies, 
Indian Institute of Science, Bangalore,India}
\author{ Agnes\ Grau}
\address{Department of Theoretical Physics, University of Granada, Granada,
Spain}
\author{ Giulia\,Pancheri} 
\address{INFN Frascati National Laboratories, Frascati, Italy}
%
%
%



\thispagestyle{empty}
\begin{flushright}                               
                                                   hep-ph/0205197  \\
                                                    IISc/CTS/04-02\\ 
\end{flushright}

\vskip 25pt

\begin{center}

{\bf Total Cross-sections\footnote{Talk given by G.Pancheri at 
Photon2001,Ascona, Switzerland}}
\vskip 25pt
  R.M. Godbole$^1$, A. Grau$^2$, and G. Pancheri$^3$ 

\bigskip
1. Centre for Theoretical Studies, Indian Institute of Science, 
Bangalore, 560012, India. E-mail: rohini@cts.iisc.ernet.in\\
\vskip 25 pt
2.   Centro Andaluz de F\'\i sica de Part\'\i culas Elementales and
   Departamento de F\'\i sica Te\'orica y del Cosmos, Universidad de
   Granada, Spain.  E-mail igrau@ugr.es\\
\vskip 25 pt
3. INFN Frascati National Laboratories, Via E. Fermi 40, I00044 
Frascati, Italy. E-mail:pancheri@lnf.infn.it\\
\bigskip
           Abstract
\end{center}
We examine the energy dependence of total cross-sections for   photon
processes and discuss the QCD contribution to the rising behaviour.
\begin{quotation}
\noindent
\end{quotation}
\newpage

\maketitle
\abstracts{
We examine the energy dependence of total cross-sections for   photon
processes and discuss the QCD contribution to the rising behaviour.}


A look at  total cross-sections\cite{plb} for the processes 
$p p, p{\bar p},  \gamma p, \gamma \gamma \rightarrow 
hadrons$  immediately raises a number of questions, like:
what gives  the energy dependence of total cross-sections? Are
photon data properly normalized? Are the predictions from 
factorization\cite{aspen}, quark counting and VMD,  consistent with the 
complete  set of data available in the same energy range?

In this talk we describe work in progress towards a QCD Description of the 
energy dependence of total cross-sections\cite{plb,bn}.
The issue has both a theoretical and a practical
interest,  as it is  necessary to have a reliable model  to predict
total hadronic cross-sections from $\gamma \gamma$ collisions, which form 
a bulk of the hadronic backgrounds at the Linear Colliders, in order that
these are properly evaluated. Indeed, convoluting the photon spectrum with
various predictions for $\gamma \gamma\rightarrow  hadrons$\cite{lc},
one finds that those  for $e^+e^-\rightarrow e^+e^-~ hadrons$ 
differ by 30$ - 40$\%. In order to reduce this uncertainty, it
is necessary to drastically reduce the range of variability  present in  
$\gamma \gamma$ collisions, where models can differ by more than a factor
two in their predictions for the total cross-section.  These differences are
due to those in the  absolute normalization and the slope with which the total 
cross-section rises in these models, all being consistent with the current 
data.

In general the task of describing 
the energy behaviour of total
cross-sections can be broken down into 
three parts:
i) the rise, ii) the initial decrease, iii)
the normalization.
The rise alone  
can be obtained 
\begin{itemize}
\item in the Regge-Pomeron  model\cite{DL},
 with $\sigma_{total}=X
s^{\epsilon}+Ys^{-\eta}$, through $s^\epsilon$, although it does not seem that
the same  power $\epsilon$ fits protons and photons\cite{adr}: one finds 
$\epsilon_{pp}=0.08,  \epsilon_{\gamma \gamma}=0.1 - 0.2$.
To overcome this problem, it has been suggested to add more power
terms, thus increasing the number of free parameters.
\item from factorization\cite{aspen},  but there remain the problem of
getting the proton-proton  cross-section from
first principles
\item
using the QCD calculable
contribution from the parton-parton cross-section, whose total
yield increases with energy\cite{halzen}
\item a combination of the above two
\end{itemize}
In the Minijet Model\cite{plb}, the rise is driven by the LO QCD 
contribution to the integrated jet cross-section
$$\sigma_{jet}=\int_{p_tmin} {{d^2\sigma_{jet}}\over{d^2{\vec p_t}}}
d^2{\vec p_t}=
\sum_{partons}\int_{p_{tmin}}d^2{\vec p_t} 
\int f(x_1)dx_1 \int f(x_2)dx_2 {{d^2\sigma^{partons}}\over{d^2{\vec p_t}}}$$
which depends on the densities
 and very dramatically on $p_{tmin}$, the minimum transverse momentum cut-off. 
To ensure unitarity, the mini-jet cross-sections are
embedded  into the eikonal formulation, which gives
the Eikonal Minijet Model in LO  QCD (EMM)

$$\sigma^{\rm inel}_{pp(\bar p)}=2\int d^2{\vec b}
[1-e^{-n(b,s)}],\ \ \ \ 
 \sigma^{\rm tot}_{pp(\bar p)}=2\int d^2{\vec b}
[1-e^{-n(b,s)/2}cos(\chi_R)]$$ 
In the EMM, one puts $\chi_R=0$. 
To proceed further, one can 
separate   the non perturbative from the
perturbative behaviour, with $n(b,s)=n_{NP}(b,s)+n_{P}(b,s)$,
and then factorize b vs. s behaviour. The simplest model has  
$n(b,s)=A(b)[\sigma_{soft}+\sigma_{jet}]$.

Taking  the matter distribution
A(b) to be the convolution of the Fourier transform of the
 form
factors of the colliding particles, the s-dependence is then entirely contained
in $\sigma_{soft}$,  parametrized so as to reproduce the low-energy
data,  and $\sigma_{jet}$, which is given by the LO QCD jet cross-sections.

\begin{figure}[htb]
\centerline{
\includegraphics*[scale=0.30]{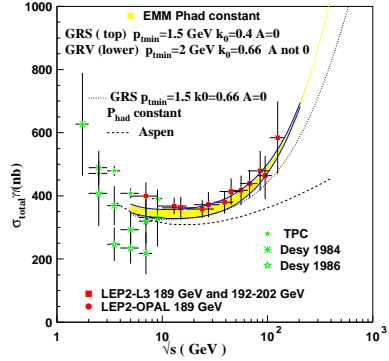}}
\vspace{-.5cm}
\caption{Predictions and data on total $\gamgam$ cross-sections.}
\label{gamgam}
\end{figure}

The consistency between   \gamp \ and \gamgam \ can be studied by applying the
EMM model with same set of parameters to the relevant data.
The total cross-section predictions for photon processes in the
EMM model include the probability $P_{had}$ 
for the photon to behave like a 
hadron, a probability expressed through a parameter obtained using
VMD, $P_{had}\approx 1/240$.
With the EMM for the \gamp\ total cross-sections, using for 
A(b) the convolution of proton (dipole) and pion-like (monopole with
scale $k_0$) formfactor, one obtains a band of values symmetrically 
encompassing  all
the data. We then apply the same formalism and the same parameters to the
\gamgam \ case, and find the band  shown in Fig.(\ref{gamgam}),
which  spans all the data, but with the lower bound slighly below the
data, especially at low energies. We also see that the Aspen model 
prediction\cite{aspen}, obtained using factorization, is clearly lower 
than the data. 
The comparison between \gamp \ and \gamgam\ indicates the existence of
a problem in the normalization of \gamgam \ data, first noticed in\cite{aspen}.
Indeed one can see that using VMD and Quark Counting to put proton and 
photon data on same scale, \gamp\ falls in place, \gamgam\ data 
remain higher than the rest,  basically the same result suggested by
the EMM model.  From these considerations, it would appear that data for 
\gamgam\  total x-section are overestimated by about 10\%. We also notice
that the normalization problem can confuse the issue of the rise. 

\begin{figure}[htb]
\centerline{
\includegraphics*[scale=0.33]{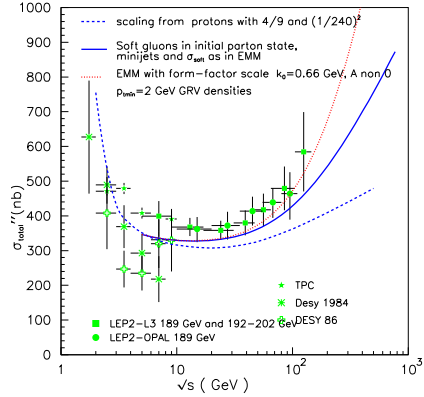}}
\caption{Effect of resummation on total cross-sections.}
\vspace{-.8cm}
\label{F4}
\end{figure}

Further refinements of the minijet model are possible, using soft gluon 
summation to include initial state acollinearity among partons. 
The model proposed\cite{bn} to do this introduced an
energy dependence in the impact parameter distribution, namely
$$n(b,s)=A_{soft}(b)\sigma_{soft}+A_{PQCD}(b,s)\sigma_{jet}^{LO}$$
with 
$A_{PQCD}(b,s)$ given by the Fourier transform of the transverse momentum
distribution of the
initial parton pair, due to initial state soft gluon radiation. Using the QCD
resummation techniques, this leads to
$$A_{PQCD}(b,s) \equiv {{e^{-h(b,s)}}\over{\int d^2{\vec b} e^{-h(b,s)}}},\
with\ 
h(b,s)=\int_{k_{min}}^{k_{max}} d^3n_{gluons}(k)[1-e^{i{\vec k}_t\cdot
{\vec b}}]$$ 
$k_{max}$, which is energy dependent,  can be taken to be the
kinematic limit, averaged over the parton   densities, while
$k_{min}=0$.
The difficulty in using $k_{min}=0$ stems from  our ignorance on $\alpha_s(k_t)$
as
 $k_t\rightarrow 0$ . To proceed further one needs to make models for this
behaviour. Our model uses a singular but integrable parametrization for
$\alpha_s$ in the infrared limit. This introduces a strong energy dependence in
the impact parameter distribution, physically understandable as  follows.
As the energy increases, one probes smaller and smaller $k_t$ values.
The more singular $\alpha_s$ is, the more is the emission of soft gluons making
the initial partons more acollinear resulting in loss of parton
luminosity and a decrease in the jet cross-section. This effect is what one 
might call  the {\it taming of the rise}.  We show in Fig.(\ref{F4}) a  
preliminary result with GRV densities and $p_{tmin}=2\ GeV$.

\end{document}